\documentstyle[aps,preprint,pra]{revtex}
\begin{document}
\newcommand{\etal}{{\em et al.}\/}
\newcommand{\IP}{inner polarization}
\newcommand{\IPF}{\IP\ function}
\newcommand{\IPFs}{\IP\ functions}
\newcommand{\erratum}[3]{\jcite{erratum}{#1}{#2}{#3}}
\newcommand{\auth}[2]{#1 #2, }
\newcommand{\oneauth}[2]{#1 #2, }
\newcommand{\jcite}[4]{#1 {\bf #2}, #3 (#4)}
\newcommand{\et}{ and }
\newcommand{\inpress}[1]{{\it #1}, in press}
\newcommand{\book}[4]{{\it #1} (#2, #3, #4)}
\newcommand{\twoauth}[4]{#1 #2 and #3 #4, }
\newcommand{\BOOK}[4]{{\it #1} (#2, #3, #4)}
\newcommand{\andauth}[2]{and #1 #2, }
\newcommand{\inbook}[5]{in {\it #1}, #2 (#3, #4, #5)}
\newcommand{\edit}[2]{Ed. #1 #2}
\newcommand{\JCP}[3]{\jcite{J. Chem. Phys.}{#1}{#2}{#3}}
\newcommand{\jms}[3]{\jcite{J. Mol. Spectrosc.}{#1}{#2}{#3}}
\newcommand{\jmsp}[3]{\jcite{J. Mol. Spectrosc.}{#1}{#2}{#3}}
\newcommand{\theochem}[3]{\jcite{J. Mol. Struct. ({\sc theochem})}{#1}{#2}{#3}}
\newcommand{\jmstr}[3]{\jcite{J. Mol. Struct.}{#1}{#2}{#3}}
\newcommand{\cpl}[3]{\jcite{Chem. Phys. Lett.}{#1}{#2}{#3}}
\newcommand{\cp}[3]{\jcite{Chem. Phys.}{#1}{#2}{#3}}
\newcommand{\pr}[3]{\jcite{Phys. Rev.}{#1}{#2}{#3}}
\newcommand{\PRA}[3]{\jcite{Phys. Rev. A}{#1}{#2}{#3}}
\newcommand{\PRL}[3]{\jcite{Phys. Rev. Lett.}{#1}{#2}{#3}}
\newcommand{\jpc}[3]{\jcite{J. Phys. Chem.}{#1}{#2}{#3}}
\newcommand{\jpca}[3]{\jcite{J. Phys. Chem. A}{#1}{#2}{#3}}
\newcommand{\jpcA}[3]{\jcite{J. Phys. Chem. A}{#1}{#2}{#3}}
\newcommand{\jcc}[3]{\jcite{J. Comput. Chem.}{#1}{#2}{#3}}
\newcommand{\molphys}[3]{\jcite{Mol. Phys.}{#1}{#2}{#3}}
\newcommand{\physrev}[3]{\jcite{Phys. Rev.}{#1}{#2}{#3}}
\newcommand{\mph}[3]{\jcite{Mol. Phys.}{#1}{#2}{#3}}
\newcommand{\cpc}[3]{\jcite{Comput. Phys. Commun.}{#1}{#2}{#3}}
\newcommand{\jcsfii}[3]{\jcite{J. Chem. Soc. Faraday Trans. II}{#1}{#2}{#3}}
\newcommand{\jacs}[3]{\jcite{J. Am. Chem. Soc.}{#1}{#2}{#3}}
\newcommand{\ijqcs}[3]{\jcite{Int. J. Quantum Chem. Symp.}{#1}{#2}{#3}}
\newcommand{\ijqc}[3]{\jcite{Int. J. Quantum Chem.}{#1}{#2}{#3}}
\newcommand{\spa}[3]{\jcite{Spectrochim. Acta A}{#1}{#2}{#3}}
\newcommand{\tca}[3]{\jcite{Theor. Chem. Acc.}{#1}{#2}{#3}}
\newcommand{\tcaold}[3]{\jcite{Theor. Chim. Acta}{#1}{#2}{#3}}
\newcommand{\jpcrd}[3]{\jcite{J. Phys. Chem. Ref. Data}{#1}{#2}{#3}}
\newcommand{\APJ}[3]{\jcite{Astrophys. J.}{#1}{#2}{#3}}
\newcommand{\astast}[3]{\jcite{Astron. Astrophys.}{#1}{#2}{#3}}
\newcommand{\arpc}[3]{\jcite{Ann. Rev. Phys. Chem.}{#1}{#2}{#3}}


\draft
\title{Towards standard methods for benchmark quality ab initio
thermochemistry --- W1 and W2 theory}
\author{Jan M.L. Martin$^*$ and Gl\^enisson de Oliveira}
\address{Department of Organic Chemistry,
Kimmelman Building, Room 262,
Weizmann Institute of Science,
76100 Re\d{h}ovot, Israel. {\em Email:} \verb|comartin@wicc.weizmann.ac.il|
}
\date{{\em J. Chem. Phys.} MS\# A9.02.198; received February 22, 1999;
revised \today}
\maketitle
\begin{abstract}
Two new schemes for computing molecular total atomization energies (TAEs)
and/or heats of formation ($\Delta H^\circ_f$) of first-and second-row
compounds to very high accuracy are presented. The more affordable
scheme, W1 (Weizmann-1) theory, yields a mean absolute
error of 0.30 kcal/mol and includes only a single, molecule-independent,
empirical parameter. It requires CCSD (coupled cluster with all
single and double substitutions) calculations in $spdf$ and $spdfg$ 
basis sets, while CCSD(T) [i.e. CCSD with a quasiperturbative treatment
of connected triple excitations] calculations are only required in
$spd$ and $spdf$ basis sets.  On workstation computers and
using conventional coupled cluster algorithms, systems as large as
benzene can be treated, while larger systems are feasible using direct
coupled cluster methods. The more rigorous scheme, W2 (Weizmann-2) theory, 
contains no empirical
parameters at all and yields a mean absolute error of 0.23 kcal/mol,
which is lowered to 0.18 kcal/mol for molecules dominated by dynamical
correlation. It involves CCSD calculations in $spdfg$ and $spdfgh$ basis
sets and CCSD(T) calculations in $spdf$ and $spdfg$ basis sets. 
On workstation computers, molecules with up to three
heavy atoms can be treated using conventional coupled cluster algorithms,
while larger systems can still be treated using a direct CCSD code.
Both schemes include corrections for scalar relativistic effects,
which are found to be vital for accurate results on second-row
compounds.
\end{abstract}

\section{Introduction}

Thermochemical data such as molecular heats of formation are among the most
crucial quantitative chemical data. Thanks to great progress made in 
recent years in both methodology and computer technology, a broad range of
empirical, semiempirical, density functional, and ab initio schemes
now exist for this purpose
(for a recent collection of reviews, see Ref.\cite{CT}).

At present, only ab initio-based methods can claim `chemical accuracy'
($\pm$1 kcal/mol). The most popular such schemes are undoubtedly the
G2\cite{g2} and G3\cite{g3} theories of Pople and coworkers (which
are based on a combination of additivity approximations and empirical
corrections applied to relatively low-level calculations), followed by
the CBS-Q\cite{cbs94,cbs96,cbs-qb3} and CBS/APNO\cite{cbs94}
methods which are intricate combinations of extrapolation and
empirical correction schemes. With the exception of CBS/APNO (which 
allows for 0.5 kcal/mol accuracy, on average\cite{ecc}, but is restricted
to first-row compounds), all these schemes allow for mean absolute errors
of about 1 kcal/mol, although errors for some individual molecules 
(e.g. SO$_2$, SiF$_4$\cite{g3}) can be {\em much} larger (e.g. about
8--12 kcal/mol for SiF$_4$ using G2 theory, and 4 kcal/mol using
G3 theory\cite{sif4}). 

In fact, many of the experimental data in the "enlarged G2 set"\cite{Pop97}
employed in the parametrization of several of these methods (notably G3
theory and several of the more recent density functional methods\cite{Bec99}) 
{\em themselves} carry experimental uncertainties larger than 
1 kcal/mol. 

The aim pursued in the present work is a more ambitious one than chemical 
accuracy. In light of the prevalent use of kJ/mol units in
the leading thermochemical tables compendia (JANAF\cite{janaf} and
CODATA\cite{Cod89}), we shall arbitrarily define a mean absolute
error of one such unit, i.e. 0.24 kcal/mol, as `calibration accuracy'
--- with the additional
constraint that no individual error be larger than the `chemical accuracy'
goal of 1 kcal/mol. 

One of us\cite{l4,c2h4tae} has recently shown that this goal is
achievable for small polyatomics using present technology.
The approach followed employed explicit treatment of inner-shell
correlation\cite{cc}, coupled cluster calculations in augmented basis
sets of $spdfg$ and $spdfgh$ quality, and extrapolation of the valence
correlation contribution to the atomization energy using formulas\cite{l4}
based on the known asymptotic convergence behavior\cite{Sch63,Hil85,Kut92}
of pair correlation energies. In this manner, total atomization energies
(TAE$_e$)
of about 15 first-row diatomics and polyatomics for which experimental
data are known to about 0.1 kcal/mol could be determined to 
within 0.25 kcal/mol on average {\em without any empirical parameters}.
(Upon introducing an empirical correction for A--N bonds, this could
be improved to 0.13 kcal/mol, clearly within the target.) In fact, using 
this method, an experimental controversy concerning the heat of formation
of gaseous boron --- a quantity that enters any ab initio or semiempirical
calculation of the heat of formation of any boron compound --- could be 
resolved\cite{bf3} by a benchmark calculation of the total atomization energy of
BF$_3$. 

Benchmark studies along similar lines by several other groups (e.g. those of 
Helgaker\cite{Hel97b}, Bauschlicher\cite{Bau98}, Dunning\cite{DunECC}) 
point in the same direction. Among those, Bauschlicher\cite{Bau98} was the 
first to 
suggest that the inclusion of scalar relativistic corrections may
in fact be essential for accurate results on second-row molecules.

High-accuracy results on second-row compounds can only be achieved 
in this manner --- as has been shown repeatedly\cite{so2,Bau95,Bau98} ---
if high-exponent $d$ and $f$ functions are added to the basis set.
As shown by one of us\cite{so2}, these `inner shell polarization functions'
address an SCF-level effect which bears little relationship to inner-shell
correlation, and actually dwarfs the latter in importance (contributions
as high as 10 kcal/mol having been reported\cite{so2,sif4}).

All these approaches carry a dual disadvantage: their extravagant
computational cost and their reliance on the quantum chemical 
expertise of the operator.

The target of the present study was to develop computational procedures
that meet the following requirements:
\begin{itemize}
\item they should have mean absolute errors on the order of 0.25 kcal/mol
or less, and problem molecules (if any) should be readily identifiable;
\item the method should be applicable to at least first-and second-row molecules;
\item it should be robust enough to be applicable in a fairly `black-box'
fashion by a nonspecialist;
\item it should rely as little as possible (preferably not at all) on 
empirical parameters, empirical additivity corrections, or other
`fudges' derived from {\em experimental} data;
\item relatedly, it should explicitly include all the physical effects that are 
liable to affect molecular binding energies of first-and second-row 
compounds, rather than rely upon absorbing them in empirical parametrization;
\item last but not least, it should be sufficiently cost-effective that
a molecule the size of, e.g., benzene should be treatable on a 
workstation computer.
\end{itemize}

In the course of this work, we will present two schemes which we shall
denote W1 and W2 (for Weizmann-1 and Weizmann-2) theories. W2 theory
yields about 0.2 kcal/mol (or better) accuracy for first-and second-row molecules
with up to four heavy atoms, and involves no empirical parameters.
W1 theory is applicable to larger systems (we shall present benzene
and trans-butadiene as examples), yet still yields a mean absolute
error of about 0.30 kcal/mol and includes only a single, molecule-independent,
empirical parameter which moreover is derived from calculated rather than
experimental results. 

\section{Computational details}

Most electronic structure calculations reported in this work were carried out
using MOLPRO 97.3\cite{m97} and MOLPRO 98.1\cite{m98} running on a Silicon 
Graphics (SGI) Octane workstation and on the SGI Origin 2000 of the 
Faculty of Chemistry. The full CCSDT (coupled cluster with all connected single,
double, and triple substitutions\cite{ccsdt}) calculations were carried out
using ACES II\cite{aces} running on a DEC Alpha 500/500 workstation. 

SCF and valence correlation calculations were carried out using 
correlation consistent\cite{Dun89,Woo93} polarized $n$-tuple zeta (cc-pV$n$Z,
or V$n$Z for short)
($n$=D, T, Q, 5, 6)
and augmented correlation consistent\cite{Ken92} polarized $n$-tuple zeta 
(aug-cc-pV$n$Z, or AV$n$Z for short) ($n$=D, T, Q, 5, 6)
basis sets of Dunning and coworkers.
The maximum angular momentum parameter $l$, which occurs in the extrapolation
formulas for the correlation energy, is identified throughout with the $n$
in V$n$Z and AV$n$Z.  Except for the calculation of the electron affinity of
hydrogen, regular V$n$Z basis sets were used throughout on hydrogen atoms. 

Most valence correlation calculations were carried out using the CCSD
(coupled cluster with all single and double substitutions\cite{Pur82}) and
CCSD(T) (i.e. CCSD followed by a quasiperturbative estimate of the effect of
connected triple excitations\cite{Rag89,Wat93}) electron correlation methods. The
CCSD(T) method is known\cite{Lee95} to be very close to an exact solution
within the given one-particle basis set if the wave function is dominated by
dynamical correlation.

Where possible, imperfections in the treatment of connected triple excitations
were estimated by comparing with full CCSDT calculations. The effect of connected
quadruple and higher excitations were estimated by small basis set
FCI (full configuration interaction)
calculations --- which represent exact solutions with a finite basis set.

Inner-shell correlation contributions were evaluated by taking the
difference between valence-only and all-electron CCSD(T) calculations in
special core-correlation basis sets. For first-row compounds, both Dunning's
ACVQZ (augmented correlation consistent core-valence quadruple zeta\cite{cvnz}) basis
set and the Martin-Taylor (MT) core correlation basis sets\cite{hf,cc} were
considered; for second-row compounds only the MT basis sets. The latter are
generated by completely decontracting a CV$n$Z or ACV$n$Z basis set, and
adding one tight $p$ function, three high-exponent $d$ functions, two
high-exponent $f$ functions, and (in the case of the MTv5z basis set) one
high-exponent $g$ function to the basis set. The additional
exponents were derived from the highest ones already present for the
respective angular momenta, successively multiplied by 3.0. The smallest
such basis set, MTvtz (based on VTZ) is also simply denoted MT.

Scalar relativistic corrections were calculated at the ACPF (averaged
coupled pair functional\cite{Gda88}) level as expectation values of the
first-order Darwin and mass-velocity terms\cite{Cow76,Mar83}. An
idea of the reliability of this approach is given by comparing a
very recent relativistic (Douglas-Kroll\cite{DK}) 
coupled cluster calculation\cite{Col98} of the relativistic
contribution to TAE[SiH$_4$], $-$0.67 kcal/mol, with the identical
value of $-$0.67 kcal/mol obtained by means of the present approach. 
For GaCl, GaCl$_2$, and GaCl$_3$ --- where relativistic effects are an order
of magnitude stronger than even in the second-row systems considered here
--- Bauschlicher\cite{GaCl3} found that differences between Douglas-Kroll
calculations and the presently followed approach amounted to 0.12 kcal/mol
or less on the binding energy. 

Spin-orbit coupling constants were evaluated at the CASSCF-CI level
using the $spdf$ part of the MTav5z basis set. (For a recent review
of the methodology involved, see Ref.\cite{Pey95}.)

Density functional calculations for the purposes of obtaining
certain reference geometries and zero-point energies
were carried out using the Gaussian 98\cite{g98} package.
Both the 
B3LYP (Becke 3-parameter\cite{B93}-Lee-Yang-Parr\cite{lyp})
and B3PW91 (Becke 3-parameter\cite{B93}-Perdew-Wang-1991\cite{pw91})
exchange-correlation functionals were considered.

Most geometry optimizations were carried out at either 
the CCSD(T)/VQZ+1 or the B3LYP/VTZ+1 (in some cases B3PW91/VTZ+1)
levels of theory, where the notation V$n$Z+1 indicates the 
addition to all second-row atoms 
of a single high-exponent $d$-type `inner polarization 
function'\cite{sio,so2} with an exponent equal to the highest
$d$ exponent in the Dunning V5Z basis set. In the past this was
found\cite{sio,so2,h2sio,so3} to recover the largest part of the
effects of inner polarization on geometries and vibrational
frequencies. (We note that for molecules consisting of first-row
atoms only, the V$n$Z+1 basis sets are equivalent to regular
V$n$Z basis sets.)

Past studies\cite{n2h2,nato,Jensen} 
of the convergence behavior of the SCF energy have shown it to be very
well described by 
a geometric extrapolation of the type first proposed by Feller\cite{Fel92},
$A+B/C^l$. Clearly, for this purpose a succession of three SCF/AV$n$Z basis
sets is required.

For the valence correlation CCSD and (T) energies, two extrapolation formulas
were considered. The first, $A+B/(l+1/2)^\alpha$, was proposed by Martin\cite{l4} 
--- the philosophy being that using the extrapolation exponent as an
adjustable parameter would enable inclusion of higher-order terms in
the asymptotic expansion
\begin{equation}
A/(L+1)^3 + B/(L+1)^4 + C/(L+1)^5 + ...
\end{equation}
while the denominator shift of 1/2 was a compromise --- for identification
of the $l$ in cc-pV$l$Z with L --- between hydrogen and nonhydrogen atoms.
The second formula, simply $A+B/l^3$, was proposed by Helgaker and
coworkers\cite{Hal98} --- where $l$ was identified with $L-1$ throughout.
Halkier et al.\cite{Hal98} already noted that in terms of the
extrapolated energy using $A+B/(l+C)^D$, the parameters C and D
were very strongly coupled, and that it only made sense to vary
one of them. 

The combination of treatments for SCF, CCSD valence correlation, (T),
imperfections in the T treatment, and connected quadruple and higher
excitations is compactly denoted here by W[p5;p4;p3;p2;p1], in which p1
denotes the basis sets involved in the SCF extrapolation, p2 the basis sets
involved in the CCSD extrapolation, p3 those in the (T) extrapolation (which
may or may not be different from p2), p4 (if present) the basis sets used in
correcting for imperfections in the treatment of connected triple
excitations, and p5 (if present) those involved in evaluating the effect of
connected quadruple and higher excitations. If any of the p's consists of a
single index, a simple additivity approximation is implied; two indices
denote a two-parameter extrapolation of the type $A+B/l^3$, while three
indices indicate a three-parameter extrapolation of the type
$A+B/(l+1/2)^\alpha$ in the case of correlation contributions, and
$A+B/C^l$ in the case of SCF contributions. For example, the level of theory
used in the previous work of Martin and Taylor would be W[TQ5;TQ5;TQ5] in
the present notation, while W[D;Q;TQ5;TQ5;TQ5] indicates
W[TQ5;TQ5;TQ5]+CCSDT/AVQZ-CCSD(T)/AVQZ+FCI/AVDZ-CCSDT/AVDZ.

\section{Atomic electron affinities as a `litmus test'}

The electron affinities of the first-and second-row atoms have often been
used as benchmarks for high-level electronic
structure methods (see e.g. the introductions to Refs.\cite{Gda99,Oli99} for reviews).
Because electron affinities involve a change in the number
of electrons correlated in the system, they are very taxing tests for any 
electron correlation method; in addition, they involve a pronounced change in the 
spatial extent of the wave function, making them very demanding in terms of the
basis set as well.

Until recently, three of the first-and second-row atomic electron 
affinities were imprecisely known experimentally (B, Al, and Si): this
situation was changed very recently by high-precision measurements for
recent experiments for B\cite{Sch98B}, Al\cite{Cal96,Sch98Al}, and Si\cite{Tho96}.

The approach we have chosen here for the SCF and valence correlation components
is summarized in our notation as W[$n$,Q,56,56,Q56] for the first-row
atoms, and W[$n$,Q,Q5,Q5,TQ5] for the second-row atoms. The effect
of inner-shell correlations was assessed at the CCSD(T)/MTav5z level, while
Darwin and mass-velocity corrections were evaluated at the ACPF/MTav5z
level. Finally, spin-orbit splittings were calculated at the CASSCF-CI
level with the $spdf$ part of a MTav5z basis set. (For technical 
reasons, the $h$ functions were omitted in both the scalar relativistic
and spin-orbit calculations, as were the $g$ functions in the latter.)

Our best computed results are compared with experiment in Table I,
where results from recent calibration studies are also summarized.

Agreement between computed and observed values can be described without reservation
as excellent: the mean absolute error amounts to 0.0009 eV.
The fact that this accuracy is obtained systematically and across the
board strongly suggest that the 'right result was obtained for the
right reason'. Upon eliminating the corrections for imperfections in CCSD(T),
i.e. restricting ourselves to W[56,56,Q56] for first-row atoms and
W[Q5,Q5,TQ5] for second-row atoms, the mean absolute error increases by
an order of magnitude to 0.009 eV, i.e. about 0.2 kcal/mol. As we shall
see below, this is essentially the type of accuracy we can obtain for 
molecules without corrections for CCSD(T) imperfections.

The importance of Darwin and mass-velocity corrections increases, as
expected, with increasing Z, and its contribution becomes quite nontrivial
for atoms like Cl. It is therefore to be expected that, e.g., in 
polar second-row molecules like ClCN or SO$_2$ 
they will contribute substantially to TAE as well.

The importance of inner-shell correlation effects is actually largest 
for Al, because of the small gap between valence and sub-valence
orbitals in the early second-row elements.

Table II compares the convergence behavior of the extrapolated valence
correlation contributions as a function of the largest basis set used,
both using the Martin three-term and Helgaker two-term formulas. While
both formulas appear to give the same answer if the underlying basis sets
are large enough, the two-term formula is by far the more stable towards 
reduction of the basis sets used in the extrapolation.
Since the use of W[Q56,Q56,Q56] is hardly an option for molecules,
the two-term formula appears to be the formula of choice.

Following the suggestion of a referee, we have considered (Table II)
the performance
of some other extrapolation formulas for the valence correlation energy.
As a point of reference, we have taken an ``experimental valence correlation
contribution to EA'', which we derived by subtracting all computed 
contributions other than the valence correlation energy from the best
experimental EAs. While some residual uncertainties may remain in some
of the individual contributions, these should be reliable to 0.001 eV on
average.

As seen in Table II, performance of the geometric series 
extrapolation\cite{Fel92} $A+B/C^n$
is outright poor: in fact, for extrapolation from AV\{D,T,Q\}Z results
the error is twice as large as that caused by not extrapolating at all.
If AV\{T,Q,5\}Z basis sets are used, mean absolute error drops to 
0.015 eV, which is still an order of magnitude larger than for the 
$A+B/l^3$ extrapolation, and only slightly better than not extrapolating 
at all. Finally, for AV\{Q,5,6\}Z basis sets, the error is three times
smaller than complete omission of extrapolation, but three times larger
than that of using any of the following formulas: $A+B/l^3$ \cite{Hal98},
$A+B/l^C$ \cite{l4}, or $A+B/(l+1/2)^4+C/(l+1/2)^6$ \cite{l4}. All three
of the latter yield a mean absolute error of about 0.001 eV, on about the
same order of accuracy as the reference values. For the smallest basis set
series AV\{D,T,Q\}Z, the mixed exponential-Gaussian
extrapolation\cite{Pet94} $A+B/\exp(l-1)+C/\exp((l-1)^2)$ represents a very
substantial improvement over $A+B/C^l$, and actually exhibits slightly better
performance than $A+B/l^3$. For the AV\{T,Q,5\}Z
series which is of greatest interest here, the Halkier et al. $A+B/l^3$
formula by far outperforms the other formulas considered. 

In short, it appears to be established that the two-term
formula of Helgaker and coworkers\cite{Hal98} is the extrapolation
method of choice overall, with the Martin three-term formulas the
second-best choice provided basis sets of AV\{T,Q,5\}Z quality are
used. The mixed exponential-Gaussian formula performs slightly
better than $A+B/l^3$ if only AV\{D,T,Q\}Z 
basis sets are used (see however Section VI.C below).

Computed spin-orbit contributions to the
electron affinities 
are compared in Table III to values obtained from
observed fine structures\cite{Moo63,Sch98B,Sch98Al}. While 
small deviations appear to persist, these may at 
least in part be due to higher-order spin-orbit effects which were neglected in
the calculation rather than to deficiencies in the electronic structure
treatment. At any rate, to the accuracy relevant for our purpose
(establishing spin-orbit corrections to molecular binding energies)
it appears to be immaterial whether the computed or the experimentally
derived values are used.

Finally, the convergence of the SCF component is so rapid that it appears to
be essentially irrelevant which extrapolation formula is used --- the 
amount bridged by the extrapolation is on the order of 0.0001 eV.

\section{Results for molecules}

Since application of electron correlation methods more elaborate than CCSD(T) 
would be well-nigh impossible for molecules of practical size, we have 
restricted ourselves to W[Q5;Q5;TQ5] and W[TQ5;TQ5;TQ5]. 

Inner-shell correlation contribution, as well as scalar relativistic corrections, 
were initially computed with the largest basis sets practicable --Ñ in most cases 
ACV5Z or MTavqz (see Table IV for details). 

From a prerelease version of a re-evaluation of the experimental data in the
G2/G3 set at the National Institute for Standards and Technology (NIST)\cite{nist}, 
we have selected 28 first-and second-row
molecules which satisfy the following criteria: (a) the uncertainty in the
experimental total atomization energy TAE is on the order of 0.25
kcal/mol or better;  (b) the molecules are not known to exhibit severe multireference
effects;  (c) anharmonic vibrational zero-point energies are available from
either experiment or high-level ab initio calculations (see footnotes to
Table IV for details).

Geometries were optimized at the CCSD(T)/VQZ+1 level, and to all second-row
atoms a complement of two tight d and one tight f function were added in
every basis set to ensure saturation in inner-shell polarization effects. In
all cases, the exponents were derived as even-tempered series
$\alpha\beta^n$ with $\beta=3.0$ and $\alpha$ the highest exponent already
present for that angular momentum.

Computed (W[Q5;Q5;TQ5]) and observed results are compared in Table IV. The
excellent agreement between theory and experiment is immediately apparent:
in many cases, the computed results fall within the already quite narrow
experimental error bars. Over the entire sample of molecules, the mean
absolute error is 0.24 kcal/mol, with the largest errors being about
0.6 kcal/mol (O$_2$ and F$_2$). 
Restricting our sample to first-row molecules only, we find a mean
absolute error of 0.24 kcal/mol, which however gets reduced to 0.17 kcal/mol
(maximum error 0.39 kcal/mol for N$_2$) upon elimination of F$_2$, NO, and
O$_2$ as having known appreciable nondynamical correlation effects. Over the
subset of 
second-row molecules in our sample MAE is 0.23
kcal/mol (maximum error 0.44 kcal/mol for H$_2$S); upon elimination of H$_2$S
and SO$_2$ this is lowered to 0.20 kcal/mol.

It should be noted that these MAEs are comparable to those found by Martin and
Taylor\cite{c2h4tae} for a sample of first-row molecules, yet unlike 
their study no correction for N-containing bonds is required here.

The possibility that the errors in F$_2$, NO, and O$_2$ are actually due to
residual basis set incompleteness and/or that the excellent agreement with
experiment for the other molecules is actually due to an error compensation
involving deficiencies in the predicted basis set limit, was examined by
carrying out W[56;56;Q56] calculations for H$_2$O, F$_2$, NO, O$_2$, N$_2$, HF, 
and CO. As seen in Table V, the predicted basis set limits do not differ
materially from their W[Q5;Q5;TQ5] counterparts, strongly suggesting that
the latter expression in fact {\em does} reach the basis set limit and that
the residual errors are largely due to imperfections in the CCSD(T) method.

While molecules liable to exhibit such errors are readily identifiable from
inspection of the largest coupled cluster amplitudes or evaluation of the
${\cal T}_1$ diagnostic\cite{t1}, an even simpler criterion is apparently offered by
the ratio TAE[SCF]/TAE[SCF+val.corr.]. In ``well-behaved'' molecules such as
CH$_4$ and H$_2$O, the SCF component makes up upwards of two-thirds of the
binding energy, while in NO and in O$_2$ it makes up no more than a third
and a fifth, respectively, of the total and F$_2$ is actually metastable
towards dissociation at the SCF level. While for some molecules of this
variety we actually obtain excellent results (e.g. ClF), this may be due to
error compensation or to the binding energies being fairly small to begin
with.

Further inspection of Table IV reveals that some of the `negligible'
contributions are in fact quite significant at the present precision level:
for instance, Darwin and mass-velocity contributions in SO$_2$ amount
to -0.71 kcal/mol
(for SiF$_4$ a somewhat extravagant -1.88 kcal/mol was found\cite{sif4}),
while atomic spin-orbit splitting in such
compounds as Cl$_2$, ClF, and SO$_2$ amounts to -1.68, -1.23, and -1.01
kcal/mol, respectively. Inner-shell correlation contributions of 2.36
(C$_2$H$_4$), 2.44 (C$_2$H$_2$), 1.68 (OCS), and 1.76 (ClCN)
kcal/mol speak for themselves; interestingly (as noted 
previously\cite{so2,h2sio}), these effects on the whole do
not seem to be more important in second-row than in first-row compounds.

Finally, we shall compare the performance of W[TQ5;TQ5;TQ5] and
W[Q5;Q5;TQ5] (Table VI). In general, the results with the three-point valence
correlation extrapolation are at best of the same quality as those with the
two-point valence correlation extrapolation and in many cases agree less
well with experiment. We therefor will use the two-point extrapolation
exclusively henceforth.

\section{W2 theory and its performance}

Having established that our 'base level of theory' can 
obtain the right results for the right reason, we shall now 
proceed to consider simplifications.

\subsection{Inner-shell correlation} 

The use of the smaller MT basis set for the scalar relativistic
contributions is found to have an effect of about 0.01 kcal/mol or less,
with 0.02 kcal/mol being the largest individual cases. This approximation
can therefore safely be made.

Using the same MT basis set for the core correlation contribution on average
affects energetics by 0.03 kcal/mol, the largest individual effects being
0.07 kcal/mol for H2S, and 0.08 kcal/mol for OCS.

Even so, in fact, the core correlation calculations are quite CPU-time
consuming, particularly for second-row compounds, due to the large number of
electrons being correlated. Any further reduction would obviously be welcome
--- it should be noted that the MT basis set was developed not with
efficiency, but with saturation (in the core-valence correlation energy) in
mind. Further experimentation revealed that the tightest p, d, and f
functions could safely be eliminated, but that further basis set reductions
adversely affect the quality of the core correlation contribution computed.
The reduced basis set shall be denoted as MTsmall, and in fact consists of a
completely decontracted cc-pVTZ basis set with two tight $d$ and one tight
$f$ functions added. Since this basis set only has about half the basis
functions of the ACVQZ basis set per heavy atom, it represents a very
significant CPU time savings (about 16 times) in a CCSD(T) calculation.
The only molecule for which we see a substantial difference with the
MT basis set is SO$_2$, for which Bauschlicher and Ricca\cite{Bau98}
previously noted that the inner-shell correlation contribution 
is unusually sensitive to the basis set.

For the evaluation of the Darwin and mass-velocity corrections, differences
with the larger MT basis set are less than 0.01 kcal/mol across the board.

A further reduction in CPU time for the core correlation 
contribution would have been achieved if MP2 or even CCSD
calculations could be substituted for their CCSD(T) counterparts. However,
as seen from Table VII, CCSD underestimates the CCSD(T) core
correlation contributions for several molecules by as much as 50\%. The
behavior of MP2 is quite similar and the MP2--CCSD differences are 
substantially smaller than the (T) contribution, suggesting that it is
the treatment of connected higher excitations that is the issue. 
Predictably, the largest (T) effects on the core correlation contribution
occur in molecules where connected triple excitations are important for
the {\em valence} binding energy as well, e.g. SO$_2$, F$_2$, Cl$_2$,
N$_2$. Conversely, in CH$_3$ or CH$_4$, which have quite small (T) 
contributions to the binding energy, CCSD does perform excellently
for the core correlation contribution. In PH$_3$ and H$_2$S, on the
other hand, substantial errors in the core correlation are seen even as
the (T) contribution to the valence correlation binding energy
is quite small --- it should be
noted, however, that both the absolute inner-shell correlation energy 
and the (T)
contribution to it are much more important in these second-row systems
than in their first-row counterparts.

One may rightly wonder whether the inner-shell contributions are in fact
converged at the CCSD(T) level. Unfortunately, if a more elaborate
treatment is already impractical for the valence correlation, this
would {\it a fortiori} be true for the inner-shell correlation. We did,
however, carry out a CCSDT/MTsmall
calculation on the N$_2$ molecule, which we chose as a representative case 
of failure of the CCSD approach for core correlation. 
The resulting CCSDT level core contribution,
0.87 kcal/mol, is only 0.05 kcal/mol larger than the CCSD(T) value of
0.82 kcal/mol, to be compared with 0.42 kcal/mol at the MP2 and 
0.52 kcal/mol at the CCSD level. It cannot be ruled out a priori that
connected quadruple and higher excitations might contribute to the
inner-shell correlation energy. However, since apparently their importance
for the valence correlation binding energy is very small (otherwise
a treatment that completely ignores them
would not yield the type of agreement with experiment found in this
work), it seems unlikely that they would greatly contribute to the
inner-shell correlation energy.

The ``G3large'' basis set used to evaluate, among other things, inner-shell
correlation effects in G3 theory\cite{g3} is still smaller than the MTsmall basis
set, and its performance therefore is certainly of interest. Alas, in Table
VII it is seen that in many cases it seriously overestimates the inner-shell
correlation energy, almost certainly because of basis set superposition
error which\cite{Bau98} is apparently more of an issue for inner-shell
correlation energies than for their valence counterparts. In G3 theory, the
inner-shell correlation is evaluated at the MP2 level: hence the two errors
cancel to a substantial extent.

\subsection{Zero-point energy}

Not in all cases is a complete anharmonic force field calculation feasible.
We find in Table VIII that B3LYP/cc-pVTZ+1 {\em harmonic} zero-point energies
scaled by 0.985 reproduce the rigorous {\em anharmonic}
zero-point energies quite nicely.
(The scaling factor is about halfway between what would be required for
fundamentals, about 0.97, and harmonics, about 1.00.\cite{azul,Sco96})

\subsection{Separate extrapolation of CCSD and (T)}

The (T) contribution makes up a relatively small part of the valence
correlation energy, while its evaluation, in large basis sets and for
systems with very many electrons, will dominate the CPU time. For
instance, in a very recent study on SiF$_4$\cite{sif4}, a 
CCSD(T) calculation with an AVQZ basis set on F and a VQZ+2d1f
basis set on Si took 50h7$'$ using
MOLPRO on an SGI Octane workstation (768 MB of memory being allocated
to the job), of which 41h30$'$ were spent in the (T) step alone.

In addition, it was previously noted by Helgaker and coworkers\cite{Hel97} that the (T)
contribution appears to converge faster with the basis set than the CCSD
correlation energy, for which reason they actually propose its separate
evaluation in a smaller basis set. In the present case, we have considered
extrapolating it from AVTZ+2d1f and AVQZ+2d1f results rather than AVQZ+2d1f
and AV5Z+2d1f, respectively. In our adopted notation, this becomes
W[TQ;Q5;TQ5].

As seen in Table VI, the difference in quality between W[TQ;Q5;TQ5] and
W[Q5;Q5;TQ5] appears to be essentially negligible.
This is an important conclusion, since it means that the largest basis set calculation to be
carried out is only at the CCSD level, at a fraction of the cost of the full
CCSD(T) counterpart --- moreover it can be done using direct algorithms.\cite{Hel96}

\subsection{Protocol for W2 theory}

The protocol obtained by introducing the succesful approximations given
above will be denoted here as W2 (Weizmann-2) theory. Its steps consist of
the following:
\begin{itemize}
\item geometry optimization at the CCSD(T)/VQZ+1 level, i.e.
CCSD(T)/VQZ if only first-row atoms are present;
\item zero-point energy obtained from a
CCSD(T)/VTZ+1 anharmonic force field or, failing that,
B3LYP/VTZ+1 frequencies scaled by 0.985 (vide infra);
\item Carry out CCSD(T)/AVTZ+2d1f and CCSD(T)/AVQZ+2d1f
single-point calculations;
\item Carry out a CCSD/AV5Z+2d1f single-point calculation;
\item the SCF component of TAE is extrapolated by
$A+B/C^l$ from SCF/AVTZ+2d1f, SCF/AVQZ+2d1f, and SCF/AV5Z+2d1f
results ($l$=3, 4, and 5, respectively);
\item the CCSD valence correlation component is obtained
from applying $A+B/l^3$ to CCSD/AVQZ+2d1f;
and CCSD/AV5Z+2d1f valence correlation energies (l=4 and 5,
respectively). It is
immaterial whether this is done on total energies or on
components to TAE;
\item the (T) valence correlation component is obtained
from applying $A+B/l^3$ results to CCSD(T)/AVTZ+2d1f
and CCSD(T)/AVQZ+2d1f values for the (T) contribution. It is
again
immaterial whether this is done on total energies or on
components to TAE;
\item core correlation computed at CCSD(T)/MTsmall level
\item scalar relativistic corrections (and, if necessary,
spin-orbit splittings) computed at ACPF/MT level. To save
CPU time, this can be combined into a single job with the
previous step.
\end{itemize}
On a typical workstation at the time of writing (e.g. an SGI Octane with 1 GB of
RAM and 2$\times$18 GB external disks) its applicability range would be about three
heavy atoms in $C_{2v}$ symmetry, although the main limiting factor would be disk
space and larger systems could be treated if a direct CCSD code were available.

We will illustrate the CPU time savings made in the W2 approach, compared to
our most rigorous calculations, using two examples: a first-row diatomic
(CO) and a second-row molecule (OCS). Using MOLPRO on an SGI Octane
workstation, the most accurate calculations reported in this work (Table IV)
required 21h36$'$ for CO and no less than 362h12$'$ for OCS. W2 theory
yields essentially identical results at a cost of 1h12$'$ (CO) or 13h42$'$
(OCS) --- a reduction by a factor of 20--30 which is typical of the
other molecules.

\section{W1 theory and its performance}

In an effort to obtain a method that is applicable to larger systems, 
we introduce a few further approximations. Relevant results can
be found in Table VI.

\subsection{Use of density functional reference geometries}

B3LYP/cc-pVTZ+1 geometries are close enough to their CCSD(T)/VQZ+1
counterparts that their use does not cause a major effect on the final
computed result. There is one notable exception to this rule for the 
molecules considered here: Cl$_2$, for which the B3LYP/VTZ+1 bond distance
of 2.0130 \AA\ is quite different from its CCSD(T)/VQZ+1 counterpart,
1.9972 \AA. (The experimental value\cite{Hub79} is 1.987$_9$ \AA.) 
While B3LYP and B3PW91 on the whole tend to produce essentially
identical geometries and harmonic frequencies\cite{Sco96}, 
it has been argued previously\cite{Gee96} that the B3PW91 functional
may be somewhat more reliable for systems with high electron
density; and in fact, the B3PW91/VTZ+1 bond distance of 
1.9912 \AA\ is much closer to the CCSD(T)/VQZ+1 and 
experimental value. 

Even so, the use of a B3LYP/VTZ+1 reference geometry still
does not affect the computed $D_e$ by more than 0.14 kcal/mol.
We conclude that CCSD(T) geometry optimizations, which will
become fairly costly for larger molecules, 
can safely be replaced by B3LYP calculations, which
can also serve for obtaining zero-point energies. 

\subsection{Further reduction of basis set sizes}

The obvious first suggestion would be to also carry out the
CCSD extrapolation using smaller basis sets, i.e. W[TQ,TQ,DTQ]. 

The effect for the SCF component of TAE is very small on condition that
the extrapolation is carried out {\em not} on the individual total energies
but rather on the computed SCF components of TAE themselves. Clearly
error compensation occurs between the molecule and the constituent
atoms.

The effect on the valence correlation component, unfortunately,
is rather more significant. Over the `training set', MAE rises
to 0.37 kcal/mol even after SO$_2$ (which clearly is a pathological 
case here) has been eliminated. Aside from the latter, eliminating
systems with mild nondynamical correlation does not lead to any
significant reduction in MAE. Also noteworthy is that, on average,
the binding energies appear to be somewhat overestimated: this
is easily explained from the fact that basis sets like AVTZ+2d1f
are not quite saturated in the {\em radial} correlation energy
either, and that therefore the TAE[AVQZ+2d1f]$-$TAE[AVTZ+2d1f]
gap will be an overestimate of the TAE[L=4]$-$TAE[L=3] gap.

\subsection{Use of empirical extrapolation exponents}

Truhlar\cite{Tru98} considered the use of $L$-extrapolation formulas 
with {\em empirical} exponents, carried out from cc-pVDZ and cc-pVTZ 
calculations, as an inexpensive alternative to very large basis set
calculations.

We will investigate here a variant of this suggestion adapted to the
present framework.
The valence correlation component to TAE will indeed be extrapolated
using the formula $A+B/l^{\beta}$, in which $\beta$ is now an empirical
parameter --- we will denote this W[Q5;Q5;TQ5]$\beta$ and the like.

We then add in all the further corrections (core correlation, scalar
relativistics, spin-orbit) that occur in W2 theory, and try to determine
$\beta$ by minimizing MAE with respect to the experimental TAE values
for our `training set'.
Not surprisingly, for W[Q5;Q5;TQ5]$\beta$ this yields an optimum exponent
($\beta$=2.98) which differs insignificantly from the `ideal' value of 3.0.
Alternatively, $\beta$ could be optimized for the best possible overlap
with the W[56;56;Q56] results: in fact, the same conclusion is obtained,
namely that making $\beta$ an empirical parameter does {\em not}
improve the quality of the results.

For W[TQ;TQ;DTQ] however, the situation is rather different. The optimum
exponent $\beta$ is found to be 3.18 if optimized against the experimental
TAE values, and 3.16 (insignificantly different) if optimized against
the W[Q5;Q5;TQ5] results. In both cases, MAE drops to 0.30 kcal/mol, and
on average no more overestimation occurs.

W[TQ;TQ;DTQ]3.18 represents a significant savings over
W2 theory. Its time-determining step in molecules with many electrons
will be the evaluation of the parenthetical triples in the AVQZ+2d1f
basis set --- their elimination would be most desirable.

A natural suggestion would then be W[DT;TQ;DTQ]$\beta$. Optimization of
$\beta$ against the experimental TAE values yields $\beta$=3.26; the not
greatly different $\beta$=3.22 is obtained by minimization of the deviation
from W[Q5;Q5;TQ5] results for the training set. Since the latter does not
explicitly depend on experimental results and therefore minor changes
in the computational protocol do not require recalculation for the
entire `training set', we will opt for the latter alternative.

In either case, we obtain MAE=0.30 kcal/mol --- for a calculation that 
requires not more than an AVTZ+2d1f basis set for the largest CCSD(T)
calculation, and an AVQZ+2d1f basis set for the largest CCSD calculation.
Again, the latter is amenable to a direct algorithm.

\subsection{Protocol for W1 theory}

We thus propose the following protocol for a 
computational level which we will call W1 (Weizmann-1)
theory:
\begin{itemize}
\item geometry optimization at the B3LYP/VTZ+1 level
(B3LYP/VTZ if only first-row atoms are present). Alternatively,
the B3PW91 exchange-correlation functional may be preferable
for some systems like Cl$_2$ --- under normal circumstances,
B3LYP/VTZ+1 and B3PW91/VTZ+1 should yield virtually identical
geometries;
\item zero-point energy obtained from 
B3LYP/VTZ+1 (or B3PW91/VTZ+1) harmonic frequencies scaled by 0.985;
\item Carry out CCSD(T)/AVDZ+2d and CCSD(T)/AVTZ+2d1f
single-point calculations;
\item Carry out a CCSD/AVQZ+2d1f single-point calculation;
\item the SCF component of TAE is extrapolated by
$A+B/C^l$ from SCF/AVDZ+2d, SCF/AVTZ+2d1f, and SCF/AVQZ+2d1f
components of TAE ($l$=2, 3, and 4, respectively)
\item set $\beta$=3.22
\item the CCSD valence correlation component is obtained
from applying $A+B/l^\beta$ to CCSD/AVTZ+2d1f
and CCSD/AVQZ+2d1f valence correlation energies (l=3 and 4,
respectively). 
In both
this and the next step, it is immaterial whether the extrapolation
is carried out on components to the total energy or to TAE;
\item the (T) valence correlation component is obtained
from applying $A+B/l^\beta$ results to CCSD(T)/AVDZ+2d
and CCSD(T)/AVTZ+2d1f values for the (T) contribution. 
\item core correlation contributions are obtained at the
CCSD(T)/MTsmall level;
\item scalar relativistic and, where necessary, spin-orbit 
coupling effects are treated at the ACPF/MTsmall level. As in
W2 theory, this latter step can be combined in a single job
with the previous step.
\end{itemize}

W1 theory can be applied to fairly large systems (see below).  CPU times
are dominated by the inner-shell correlation contribution (particularly 
for second-row compounds), which is reflected in the relatively small 
time reduction compared to W2 theory --- e.g., from 1h12$'$ to 24$'$ for
CO and from 13h42$'$ to 8h48$'$ for OCS. In addition --- contrary to
W2 theory --- W1 theory exhibits a pronounced difference in performance
between first-row and second-row compounds: for the species in Table VI,
MAE is 0.26 kcal/mol for first-row, but 0.40 kcal/mol for second-row
compounds. Since the CPU time gap between W1 and W2 theory is fairly
narrow for second-row species, we conclude that for accurate work on
second-row species --- unless precluded by disk space or memory
limitations --- it may well be worth to `walk the extra mile' and
carry out a W2 rather than a W1 calculation. For first-row systems,
on the contrary, W1 may well seem the more attractive of the two.

\section{Sample applications to larger systems}

By way of illustration, we have carried out some W1 theory
calculations on trans-1,3-butadiene
and benzene.
All relevant computed and observed results are summarized in Table IX.
The example of benzene is representative and will be 
discussed here in detail --- it should be mentioned that
the calculation was carried out in its entirety
on an SGI Octane workstation with 2x18 GB external
SCSI disks.

The reference geometry was obtained at the B3LYP/cc-pVTZ level\cite{dft}.
The zero-point energy at that level, after scaling by 0.985, is found to
be 62.04 kcal/mol. 

The SCF component of TAE is predicted to be 1044.95 kcal/mol
at the one-particle basis set limit, of which only 0.39
kcal/mol is covered by the geometric extrapolation.
Of the CCSD valence correlation component of 291.07
kcal/mol, however, some 10.11 kcal/mol is covered by 
the extrapolation, which also accounts for 2.13 kcal/mol
out of the 26.55 kcal/mol connected triple excitations
contribution.

The inner-shell correlation contribution is quite sizable
at 7.09 kcal/mol, although this number is not qualitatively
different from that for three acetylenes or three ethylenes.
Finally, Darwin and mass-velocity terms contribute a small
but significant -0.96 kcal/mol, and atomic spin-orbit
splitting another -0.51 kcal/mol. All adds up to 1367.95
kcal/mol at the bottom of the well, or 1305.92 kcal/mol
at 0 K, which is in excellent agreement with the experimental
value of  1306.1$\pm$0.12 kcal/mol from the NIST WebBook\cite{WebBook}.

The CCSD/VQZ calculation took 10h10' with MOLPRO on the Octane,
the CCSD(T)/VTZ calculation 1h48' on a single CPU on the Origin 2000.
By far the most time-consuming part of the calculation was the inner-shell
correlation contribution, at 67h46', to which another 4h52' should be
added for the Darwin and mass-velocity contribution.
We see similar trends in the results for trans-butadiene,
which agree with experiment to virtually within the 
stated experimental uncertainty; for allene, we obtain
a value intermediate between the two experimental 
values proposed in the WebBook.

We find for both molecules that the sum of
core-correlation and relativistic contributions can be
quite well estimated by additivity approximations. For
instance, the core correlation and scalar relativistic
contributions with the same basis set for C$_2$H$_4$ are
+2.360 and -0.330 kcal/mol, respectively, adding up
to 2.030 kcal/mol. Assuming 2 and 3 times this
`C=C bond equivalent' for butadiene and
benzene, respectively, yields estimated contributions 
of 4.06 (trans-butadiene), and 6.09 (benzene) kcal/mol, which 
agree excellently with the directly computed values
of 4.02 and 6.12 kcal/mol, respectively. Considering
that inner-shell correlation effects should be fairly
local in character, such schemes should work quite well
for larger organic systems where the valence calculation
would still be feasible but the explicit inner-shell
calculation would not be.
 
\section{Conclusions}

We have developed and presented two quasi-`black box' schemes for
high-accuracy calculation of molecular atomization energies or, equivalently,
molecular heats of formation, of first-and second-row compounds.

The less expensive scheme, W1 (Weizmann-1) theory, yields a mean absolute
error of 0.30 kcal/mol and includes only a single, molecule-independent,
empirical parameter. It requires no larger-scale calculations than
CCSD/AVQZ+2d1f and CCSD(T)/AVTZ+2d1f (or, for nonpolar first-row 
compounds, CCSD/VQZ and CCSD(T)/VTZ). On workstation computers and
using conventional coupled cluster algorithms, systems as large as
benzene can be treated, while larger systems are feasible using direct
coupled cluster methods.

The more expensive scheme, W2 (Weizmann-2) theory, contains no empirical
parameters at all and yields a mean absolute error of 0.23 kcal/mol,
which is lowered to 0.18 kcal/mol for molecules dominated by dynamical 
correlation. On workstation computers, molecules with up to three
heavy atoms can be treated using conventional coupled cluster algorithms,
while larger systems can still be treated using a direct CCSD code.

The inclusion of scalar relativistic (Darwin and mass-velocity)
corrections is essential for good results in second-row
compounds, particularly highly polar ones. Inclusion of inner-shell
correlation contributions is absolutely essential: the basis set denoted as
MTsmall (for Martin-Taylor small) appears to represent the best compromise 
between quality and computational expense. We do not recommend the use of 
lower-level electron correlation methods than CCSD(T) for the 
evaluation of the inner-shell contribution.

Among the several infinite-basis set extrapolation formulas for the 
correlation energy examined, the three-parameter $A+B/(l+1/2)^\alpha$
expression proposed by Martin\cite{l4} 
and the $A+B/l^3$ expression proposed
by Helgaker and coworkers\cite{Hal98} yield the best results for sufficiently
large basis sets, with the latter formula to be preferred on
grounds of stability of the extrapolated results with the basis sets used.
Geometric and mixed geometric-Gaussian extrapolation formulas\cite{Fel92,Pet94} 
are unsatisfactory when applied to the correlation energy, 
although they appear to be appropriate
for the SCF component.

The main limiting factor for the quality of our calculations at this stage 
appears to be imperfections in the CCSD(T) method. This assertion is
supported by the fact that the mean absolute error in the computed electron
affinities of the atoms H, B--F and Al--Cl drops from 0.009 eV to 0.0009 eV
if CCSDT and full CI corrections are included.

Extrapolation of the (T) contribution to the correlation energy can, at no
loss in accuracy, be carried out using smaller basis sets than the CCSD
contribution.

\acknowledgments

JM is a Yigal Allon Fellow, the incumbent of the Helen and Milton
A. Kimmelman Career Development Chair (Weizmann Institute), and
an Honorary Research Associate (``Onderzoeksleider
in eremandaat'') of the
National Science Foundation of Belgium (NFWO/FNRS). GdO acknowledges
the Feinberg Graduate School (Weizmann Institute) 
for a Postdoctoral Fellowship.
This research was supported by the Minerva Foundation, Munich,
Germany.
The authors acknowledge enlightening discussions with 
(in alphabetical order) Drs. C. W. Bauschlicher Jr. (NASA Ames
Research Center), Dr. Thom H. Dunning (PNNL), Prof. Trygve Helgaker
(U. of Oslo, Norway), Dr. Frank Jensen (Odense U., Denmark), 
Dr. Timothy J. Lee (NASA Ames Research Center), and
Prof. Peter R. Taylor (UCSD and National Partnership for Advanced
Computing Infrastructure), and thank Dr. Peter Stern for technical
assistance with various computer systems.


\begin{thebibliography}{99}


\bibitem{CT} {\it Computational Thermochemistry}
(Irikura, K. K. and Frurip, D. J., Eds.), ACS Symposium Series, Nr. 677, American
Chemical Society, Washington, DC, 1998.

\bibitem{g2} \auth{L. A.}{Curtiss} \auth{K.}{Raghavachari}
\auth{G. W.}{Trucks} \andauth{J. A.}{Pople}
\JCP{94}{7221}{1991}

\bibitem{g3} \auth{L. A.}{Curtiss} \auth{K.}{Raghavachari} 
\auth{P. C.}{Redfern} \auth{V.}{Rassolov} \andauth{J. A.}{Pople} \JCP{109}{7764}{1998}


\bibitem{cbs94} \auth{J. A.}{Montgomery Jr.} \auth{J. W.}{Ochterski}
\andauth{G. A.}{Petersson}
\JCP{101}{5900}{1994}
and references therein

\bibitem{cbs96} \auth{J. W.}{Ochterski} \auth{G. A.}{Petersson} \andauth{J. A.}{Montgomery Jr.}
\JCP{104}{2598}{1996}

\bibitem{cbs-qb3} \auth{J. A.}{Montgomery, Jr.} \auth{M. J.}{Frisch}
\auth{J. W.}{Ochterski} \andauth{G. A.}{Petersson} \JCP{110}{2822}{1999}

\bibitem{ecc} J. M. L. Martin, {\it Benchmark studies on small molecules},
in {\it Encyclopedia of Computational Chemistry}, 5 vols. (Wiley, NY, 1998)

\bibitem{sif4} J. M. L. Martin and P. R. Taylor, {\it J. Phys. Chem. A}
{\bf 103}, 4427 (1999).

\bibitem{Pop97} \auth{L. A.}{Curtiss} \auth{K.}{Raghavachari} \auth{P.
C.}{Redfern} \andauth{J. A.}{Pople} \JCP{106}{1063}{1997}

\bibitem{Bec99} A. D. Becke, \jcc{20}{63}{1999}
and references therein.; see also
A. D. Becke, 
\JCP{107}{8554}{1997};
H. L. Schmider and A. D. Becke,
\JCP{108}{9624}{1998};
ibid. {\bf 109}, 8188 (1998).

\bibitem{janaf}
{\it NIST-JANAF Thermochemical Tables, 4th Edition}, Ed. M. W. Chase Jr.,
{\it J. Phys. Chem. Ref. Data} Monograph 9 (1998).

\bibitem{Cod89} \auth{J.D.}{Cox} \auth{D.D.}{Wagman} \andauth{V.A.}{Medvedev}
\BOOK{CODATA key values for thermodynamics}{Hemisphere}{New York}{1989}



\bibitem{l4} \auth{J. M. L.}{Martin}
\cpl{259}{669}{1996}

\bibitem{c2h4tae} \twoauth{J. M. L.}{Martin}{P. R.}{Taylor}
\JCP{106}{8620}{1997}

\bibitem{cc} \oneauth{J. M. L.}{Martin} \cpl{242}{343}{1995}

\bibitem{Sch63} \oneauth{C.}{Schwartz} 
\inbook{Methods in Computational
Physics 2}{\edit{B. J.}{Alder}}{Academic Press}{New York}{1963}.

\bibitem{Hil85} \oneauth{R. N.}{Hill} \JCP{83}{1173}{1985}

\bibitem{Kut92} W. Kutzelnigg and J. D. Morgan III, \JCP{96}{4484}{1992};
\erratum{97}{8821}{1992}

\bibitem{bf3} \twoauth{J. M. L.}{Martin}{P. R.}{Taylor} \jpca{102}{2995}{1998}

\bibitem{Hel97b} 
\auth{T.}{Helgaker} \auth{W.}{Klopper} \auth{H.}{Koch} \andauth{J.}{Noga}
\JCP{106}{9639}{1997}

\bibitem{Bau98} \twoauth{C. W.}{Bauschlicher Jr.}{A.}{Ricca}
\jpca{102}{8044}{1998}

\bibitem{DunECC} T. H. Dunning Jr., "Basis sets: correlation consistent" in 
{\it Encyclopedia of Computational Chemistry}, 5 vols. (Wiley, NY, 1998); see
also 
\twoauth{D.}{Feller}{K. A.}{Peterson} \JCP{108}{154}{1998}

\bibitem{so2}
\oneauth{J. M. L.}{Martin} \JCP{108}{2791}{1998}

\bibitem{Bau95} \twoauth{C. W.}{Bauschlicher Jr.}{H.}{Partridge}
\cpl{240}{533}{1995}

\bibitem{m97} \auth{H.-J.}{Werner} \andauth{P. J.}{Knowles}
MOLPRO 97.3, a package of {\em ab initio} programs,
with contributions from
\auth{J.}{Alml\"of} \auth{R. D.}{Amos} \auth{A.}{Berning} \auth{D. L.}{Cooper}
\auth{M. J. O.}{Deegan} \auth{A. J.}{Dobbyn}
\auth{F.}{Eckert} \auth{S. T.}{Elbert} \auth{C.}{Hampel} \auth{R.}{Lindh}
\auth{A. W.}{Lloyd} \auth{W.}{Meyer} \auth{A.}{Nicklass}
\auth{K. A.}{Peterson} \auth{R. M.}{Pitzer} \auth{A. J.}{Stone}
\auth{P. R.}{Taylor} \auth{M. E.}{Mura} \auth{P.}{Pulay}
\auth{M.}{Sch\"utz} \auth{H.}{Stoll} Thorsteinsson, T.

\bibitem{m98} 
MOLPRO is a package of ab initio programs written by H.-J. Werner and P. J. Knowles, with contributions from J. Alml{\"o}f,
R. D. Amos, A. Berning, D. L. Cooper, M. J. O. Deegan, A. J. Dobbyn, F. Eckert, S. T. Elbert, C. Hampel, R. Lindh,
A. W. Lloyd, W. Meyer, A. Nicklass, K. Peterson, R. Pitzer, A. J. Stone, P. R. Taylor, M. E. Mura, P. Pulay, M. Sch\"utz,
H. Stoll, and T. Thorsteinsson.

\bibitem{ccsdt} \twoauth{J.}{Noga}{R. J.}{Bartlett} \JCP{86}{7041}{1987}
\erratum{89}{3401}{1988}

\bibitem{aces} \auth{J. F.}{Stanton} \auth{J.}{Gauss}
\auth{J. D.}{Watts} \auth{W.}{Lauderdale} \andauth{R. J.}{Bartlett}
(1996) ACES II, an ab initio program system, incorporating the
MOLECULE vectorized molecular integral program by Alml\"of, J., and Taylor, P. R.,
and a modified version of the ABACUS integral derivative package by
\auth{T.}{Helgaker} \auth{H. J. Aa.}{Jensen} \auth{P.}{J{\o}rgensen}
\auth{J.}{Olsen} \andauth{P. R.}{Taylor}.

\bibitem{Dun89}  \auth{T. H.}{Dunning Jr.}
\JCP{90}{1007}{1989}

\bibitem{Woo93} \twoauth{D. E.}{Woon}{T. H.}{Dunning Jr.}
\JCP{98}{1358}{1993}.

\bibitem{Ken92} \auth{R. A.}{Kendall} \auth{T. H.}{Dunning} \andauth{R.
J.}{Harrison}
\JCP{96}{6796}{1992}

\bibitem{Pur82} \twoauth{G. D.}{Purvis III}{R. J.}{Bartlett} \JCP{76}{1910}{1982}

\bibitem{Rag89}
\auth{K.}{Raghavachari} \auth{G. W.}{Trucks} \auth{J. A.}{Pople}
\andauth{M.}{Head-Gordon} {\it Chem. Phys. Lett.} {\bf 157}, 479 (1989)

\bibitem{Wat93} \auth{J. D.}{Watts} \auth{J.}{Gauss} \andauth{R. J.}{Bartlett}
\JCP{98}{8718}{1993}

\bibitem{Lee95} \twoauth{T. J.}{Lee}{G. E.}{Scuseria}
\inbook{Quantum mechanical
electronic structure calculations with chemical accuracy}{\edit{S.
R.}{Langhoff}}{Kluwer}{Dordrecht, The Netherlands}{1995}, pp. 47--108.

\bibitem{cvnz} \twoauth{D. E.}{Woon}{T. H.}{Dunning Jr.} \JCP{103}{4572}{1995}

\bibitem{hf} \twoauth{J. M. L.}{Martin}{P. R.}{Taylor}
\cpl{225}{473}{1994}

\bibitem{Gda88} \twoauth{R. J.}{Gdanitz}{R.}{Ahlrichs} \cpl{143}{413}{1988}

\bibitem{Cow76} \twoauth{R. D.}{Cowan}{M.}{Griffin}
 \jcite{J. Opt. Soc. Am.}{66}{1010}{1976}

\bibitem{Mar83} \oneauth{R. L.}{Martin} \jpc{87}{750}{1983}

\bibitem{DK}
\twoauth{M.}{Douglas}{N. M.}{Kroll} \jcite{Ann. Phys. (NY)}{82}{89}{1974};
\auth{R.}{Samzow} \auth{B. A.}{He\ss} \andauth{G.}{Jansen}
\JCP{96}{1227}{1992}
and references therein.

\bibitem{Col98} \twoauth{C. L.}{Collins}{R. S.}{Grev}
\JCP{108}{5465}{1998}

\bibitem{GaCl3} \oneauth{C. W.}{Bauschlicher Jr.} \jcite{Theor. Chem. Acc.}{101}{421}{1999}

\bibitem{Pey95} \auth{B. A.}{He{\ss}} \auth{C. M.}{Marian}
\andauth{S. D.}{Peyerimhoff} 
in {\it Modern Electronic Structure Theory, Vol. 1}, (Yarkony, D. R., Ed.);
World Scientific, Singapore, {\bf 1995}, p. 152--278.

\bibitem{g98}
\auth{M. J.}{Frisch} \auth{G. W.}{Trucks} \auth{H. B.}{Schlegel}
\auth{G. E.}{Scuseria} \auth{M. A.}{Robb} \auth{J. R.}{Cheeseman}
\auth{V. G.}{Zakrzewski} \auth{J. A.}{Montgomery} \auth{R. E.}{Stratmann}
\auth{J. C.}{Burant} \auth{S.}{Dapprich} \auth{J. M.}{Millam}
\auth{A. D.}{Daniels} \auth{K. N.}{Kudin} \auth{M. C.}{Strain}
\auth{O.}{Farkas} \auth{J.}{Tomasi} \auth{V.}{Barone} \auth{M.}{Cossi}
\auth{R.}{Cammi} \auth{B.}{Mennucci} \auth{C.}{Pomelli} \auth{C.}{Adamo}
\auth{S.}{Clifford} \auth{J.}{Ochterski} \auth{G. A.}{Petersson}
\auth{P. Y.}{Ayala} \auth{Q.}{Cui} \auth{K.}{Morokuma} \auth{D. K.}{Malick}
\auth{A. D.}{Rabuck} \auth{K.}{Raghavachari} \auth{J. B.}{Foresman}
\auth{J.}{Cioslowski} \auth{J. V.}{Ortiz} \auth{B. B.}{Stefanov} \auth{G.}{Liu}
\auth{A.}{Liashenko} \auth{P.}{Piskorz} \auth{I.}{Komaromi} \auth{R.}{Gomperts}
\auth{R. L.}{Martin} \auth{D. J.}{Fox} \auth{T.}{Keith} \auth{M. A.}{Al-Laham}
\auth{C. Y.}{Peng} \auth{A.}{Nanayakkara} \auth{C.}{Gonzalez}
\auth{M.}{Challacombe} \auth{P. M. W.}{Gill} \auth{B. G.}{Johnson}
\auth{W.}{Chen} \auth{M. W.}{Wong} \auth{J. L.}{Andres}
\auth{M.}{Head-Gordon} \auth{E. S.}{Replogle} \andauth{J. A.}{Pople}
{\it Gaussian 98, Revision A.3} (Gaussian, Inc., Pittsburgh, PA, 1998).

\bibitem{B93}
A. D. Becke, \JCP{98}{5648}{1993}.

\bibitem{lyp} C. Lee, W. Yang, and R. G. Parr, \jcite{Phys. Rev. B}{37}{785}{1988}

\bibitem{pw91}
\auth{J. P.}{Perdew} \auth{J. A.}{Chevary}
\auth{S. H.}{Vosko} \auth{K. A.}{Jackson}
\auth{M. R.}{Pederson} \auth{D. J.}{Singh} \andauth{C.}{Fiolhais}
\jcite{Phys. Rev. B}{46}{6671}{1992} and references therein

\bibitem{sio} \twoauth{J. M. L.}{Martin}{O.}{Uzan}
\cpl{282}{16}{1998}

\bibitem{h2sio}
\oneauth{J. M. L.}{Martin} \jpcA{102}{1394}{1998}

\bibitem{so3} J. M. L. Martin, \jcite{Spectrochim. Acta A}{55}{713}{1999}

\bibitem{n2h2} J. M. L. Martin and P. R. Taylor, Mol. Phys. {\bf 96}, 681 (1999)

\bibitem{nato}
\auth{J. M. L.}{Martin}
in NATO ASI Symposium Volume {\it Energetics of stable
molecules and reactive intermediates} (ed. M. E. Minas da
Piedade), NATO ASI Series C {\bf 535} (Kluwer, Dordrecht, 1999), pp. 373-415.

\bibitem{Jensen} F. Jensen, J. Chem. Phys., submitted.

\bibitem{Fel92} \oneauth{D.}{Feller} \JCP{96}{6104}{1992}

\bibitem{Hal98}\auth{A.}{Halkier} \auth{T.}{Helgaker} 
\auth{P.}{J{\o}rgensen} \auth{W.}{Klopper}
\auth{H.}{Koch} \auth{J.}{Olsen} \andauth{A. K.}{Wilson} \cpl{286}{243}{1998}

\bibitem{Gda99} R. J. Gdanitz, \JCP{110}{706}{1999}

\bibitem{Oli99} G. de Oliveira, J. M. L. Martin, F. De Proft, 
and P. Geerlings, Phys. Rev. A, in press.

\bibitem{Sch98B} M. Scheer, R. C. Bilodeau, and H. K. Haugen,
\PRL{80}{2562}{1998}

\bibitem{Cal96} D. Calabrese, A.M. Covington, and J.S. Thompson, Phys. Rev. A
54, 2797 (1996).

\bibitem{Sch98Al} M. Scheer, R.C. Bilodeau, J. Thogersen, and H.K. Haugen,
Phys. Rev. A 57, R1493 (1998).

\bibitem{Tho96} J. Thogersen, L.D. Steele, M. Scheer, C.A. Brodie, and H.K.
Haugen, J. of Phys. B, 29, 1323 (1996).

\bibitem{Pet94} K. A. Peterson, D. E. Woon, and T. H. Dunning Jr., 
\JCP{100}{7410}{1994}.

\bibitem{Moo63} \auth{C. E.}{Moore} {\it Atomic energy levels},
Natl. Bur. Stand. (US) Circ. {\bf 1949}, {\it 467}.

\bibitem{nist} R. D. Johnson III and J. F. Liebman, personal communication.

\bibitem{t1} \twoauth{T. J.}{Lee}{P. R.}{Taylor}
\ijqcs{23}{199}{1989}

\bibitem{azul} J. M. L. Martin, J. El-Yazal, and J. P. Fran\c{c}ois,
\jpc{100}{15358}{1996}

\bibitem{Sco96} \twoauth{A. P.}{Scott}{L.}{Radom} \jpc{100}{16502}{1996}

\bibitem{Hel97}
\auth{W.}{Klopper} \auth{J.}{Noga} \auth{H.}{Koch} 
\andauth{T.}{Helgaker} \tca{97}{164}{1997}

\bibitem{Hel96} 
e.g. \auth{H.}{Koch} \auth{A.}{Sanchez de Meras} \auth{T.}{Helgaker}
\andauth{O.}{Christiansen} \JCP{104}{4157}{1996}

\bibitem{Hub79} \twoauth{K. P.}{Huber}{G.}{Herzberg}
{\it Constants of Diatomic Molecules} (Van Nostrand Reinhold, 
New York, 1979)

\bibitem{Gee96}
P. Geerlings, F. De Proft, and J. M. L. Martin, 
in {\it Theoretical and Computational
Chemistry, Vol. 4: Recent developments and applications of 
modern density functional theory} 
(ed. J. Seminario), Elsevier, New York, 1996, pp. 773. 


\bibitem{Tru98} \oneauth{D. G.}{Truhlar} \cpl{294}{45}{1998}

\bibitem{dft} J. M. L. Martin, J. El-Yazal, and J. P. Fran\c{c}ois, 
\molphys{86}{1437}{1995}.

\bibitem{WebBook}
H.Y. Afeefy, J.F. Liebman, and S.E. Stein, "Neutral Thermochemical Data" in 
NIST Chemistry WebBook, NIST Standard Reference Database Number 69, 
Eds. W.G. Mallard and P.J. Linstrom, November 1998, 
National Institute of Standards and Technology, Gaithersburg MD, 20899 
(http://webbook.nist.gov). 





























\end{thebibliography}
\end{document}